\documentclass[12pt,preprint]{aastex}

\shorttitle{Searching for $z\gtrsim 6$ Quasars}

\begin{document}
\title {\bf An Exploratory Search for $z\gtrsim 6$ Quasars in the UKIDSS Early Data Release\altaffilmark{1}}

\author{Eilat Glikman\altaffilmark{2}, Alexander Eigenbrod\altaffilmark{3}, S.~G. Djorgovski\altaffilmark{2,3}, Georges Meylan\altaffilmark{3}, David  Thompson\altaffilmark{4}, Ashish Mahabal\altaffilmark{2}, Fr\'{e}d\'{e}ric Courbin\altaffilmark{3}}

\altaffiltext{1}{The data presented herein were obtained at the W.M. Keck
Observatory, which is operated as a scientific partnership among the
California Institute of Technology, the University of California and the
National Aeronautics and Space Administration. The Observatory was made
possible by the generous financial support of the W.M. Keck Foundation.}

\altaffiltext{2}{Astronomy Department, California Institute of Technology,
Pasadena, CA  91125, email: [eilatg,george,aam]@astro.caltech.edu}

\altaffiltext{3}{Laboratoire d'Astrophysique, Ecole Polytechnique F\'{e}d\'{e}rale de Lausanne (EPFL), CH-1290 Sauverny, Switzerland.}

\altaffiltext{4}{Large Binocular Telecope Observatory, University of Arizona, Tucson, AZ 85721, email: dthompson@as.arizona.edu }

\begin{abstract}
We conducted an exploratory search for quasars at $z\sim 6 - 8$,  using the Early Data Release from United Kingdom Infrared Deep Sky survey (UKIDSS) cross-matched to panoramic optical imagery.  High redshift quasar candidates are chosen using multi-color selection in $i,z,Y,J,H$ and $K$ bands.  After removal of apparent instrumental artifacts, our candidate list consisted of 34 objects.  We further refined this list with deeper imaging in the optical for ten of our candidates. Twenty-five candidates were followed up spectroscopically in the near-infrared and in the optical. We confirmed twenty-five of our spectra as very low-mass main-sequence stars or brown dwarfs, which were indeed expected as the main contaminants of this exploratory search.  The lack of quasar detection is not surprising: the estimated probability of finding a single $z>6$ quasar down to the limit of UKIDSS in the 27.3 square degrees of the EDR is $<5\%$. We find that the most important limiting factor in this work is the depth of the available optical data.  Experience gained in this pilot project can help refine high-redshift quasar selection criteria for subsequent UKIDSS data releases.
\end{abstract}


\section {Introduction}
Quasars remain one of the most powerful probes of the early universe, providing measurements and constraints for reionization, early structure formation, and early chemical enrichment.  Absorption spectra of quasars at $z > 5.5$ provide some of the key measurements for the evolution of the inter-galactic medium (IGM) transmission at the end of the cosmic reionization \citep[see reviews by, e.g.][and references therein]{Fan06b,Djorgovski06,Loeb06}.  This includes direct measurements of the optical depth of IGM absorption, and estimates of the radii of quasar Stromgren spheres, which constrain the neutral hydrogen fraction in the regimes $x_\mathrm{HI} \sim 10^{-4} - 10^{-2}$, and $\sim 10^{-1}$, respectively.  Metallic line absorbers seen in their spectra constrain the early chemical evolution of the IGM.  Abundance analysis of quasar spectra reaching out to $z \sim 6$ implies a high degree of chemical evolution in their hosts, with metallicities up to $\sim 10$ Z$_\sun$ and Fe/Mg ratios indicative of a very early chemical enrichment by type I SNe \citep[e.g.][]{Dietrich03a,Dietrich03b}.   They seem to be powered by black holes with masses reaching a few times $10^9$ M$_\sun$ \citep[e.g.,][]{Vestergaard04}, implying an early and extremely efficient mass assembly. Luminous quasars may mark some of the densest and most biased spots in the early universe, which may have enhanced rates of galaxy formation \citep[and references therein]{Stiavelli05,Djorgovski05}.

 All of these observations represent non-trivial constraints for theoretical models, which become sharper as the quasar redshift  increases.  The problem in pushing to higher redshifts is that the Ly$\alpha$ line and the associated flux drop used to color-select quasars move outside the CCD wavelength range for $z > 6.5$ or so.  Since the most luminous quasars are extremely rare, especially at such high redshifts, a large, moderately deep, panoramic IR sky survey is needed.

The UK Infrared Telescope (UKIRT) Infrared Deep Sky Survey \citep[UKIDSS,][]{Lawrence06} offers an unprecedented opportunity to discover bright quasars at redshifts beyond the $z \sim 6.5$ barrier which limits the purely optical surveys such as the Sloan Digital Sky survey \citep[SDSS,][]{York00} and Palomar Quest (PQ)\footnote{\url{http://palquest.org}}.  The luminosity function of \citep{Fan01} predicts that about 10 such quasars should be discovered over the 4000 deg$^2$, the final coverage area planned for the UKIDSS Large Area Survey (LAS). 

With this newly available data, we have embarked upon an exploratory study, using catalogs from the UKIDSS Early Data Release (EDR) to conduct a preliminary search for high-redshift quasars.  \citet{Venemans07} took advantage of the overlap between the UKIDSS first data release (DR1) and the SDSS deep survey \citep[Stripe 82;][]{AM07} which reaches $\sim 2$ magnitudes deeper than the single-scan SDSS images.  Using optical-to-near-infrared color selection, \citet{Venemans07} found ULAS J020332.38$+$001229.2, a $z=5.86$ quasar.  This illustrates the potential of UKIDSS for finding high redshift quasars.

The goal of this study is to explore this further, with the available UKIDSS EDR data, combined with available optical data from the SDSS and PQ surveys.  Our purpose here is to understand better the mothodology, the limitations of the available data, and the practical issues posed by the selection of high-redshift quasar candidates in this regime.  We do not try to estimate the quasar luminosity function at these previously unprobed redshifts, since the initial available UKIDSS EDR coverage is too small to expect detections of even a few $z>6.5$ quasars, as we discuss below. 

In \S2 we describe our selection technique, including our reliance on SDSS and PQ imaging in \S2.1 and \S2.2, respectively.  In \S3 we describe our followup observations; we present the imaging data in \S3.1 and our spectroscopy in \S3.2. In \S 4 we analyze our spectra and in \S 5 we provide a brief discussion of our results.

\section {Selection Technique}

The UKIDSS survey is a near-infrared imaging sky survey which will cover 7500 deg$^2$ at its completion.  The survey is comprised of five separate surveys with varying depths, coverage areas, and filter combinations aimed at specific science goals.  The Large Area Survey has the ideal design for high redshift quasar searches.  Covering $\sim$4000 deg$^2$ in the $Y$, $J$, $H$ and $K$ passbands and designed to reach signal-to-noise ratio of 5 at $K=18.1$ (Vega magnitudes), the survey is primed to detect high redshift ($z\sim 7$) quasars.

The UKIDSS Early Data Release \citep[EDR,][]{Dye06} covers 27.3 deg$^2$ of LAS data. The median 5$\sigma$ point source depths for this data in the four LAS filters are $Y=20.23$, $J=19.52$, $H=18.73$ and $K=18.06$ (Vega magnitudes).  We select our high redshift quasar candidates by combining near-infrared data from the LAS in the EDR of UKIDSS with optical data from SDSS and PQ and selecting objects in color-color space based on the synthetic colors for quasars presented in \citet{Hewett06}.

Using the WFCAM Science Archive (WSA) SQL interface\footnote{\url{http://surveys.roe.ac.uk/wsa/edrplus\_release.html}}  we selected the entire EDR catalog by querying for all objects within a set of cones chosen to include the entire LAS area as shown in Figures 9 and 10 of \citet{Dye06}.  We select all objects within a 90 arcminute radius centered on a series of 17 pointings in the LAS2 and LAS4 regions.  In LAS2 we queried for objects in cones centered on $14^h52^m30^s$ $+$06\degr30\arcmin, $15^h29^m$ $+$05\degr 40\arcmin, $15^h39^m$ $+$06\degr 00\arcmin, and $15^h49^m$ $+$05\degr 40\arcmin. In the LAS4 area, we centered our queries on 13 equatorial positions (Dec.= 00\degr00\arcmin) with R.A.= $12^h13^m$, $12^h24^m$, $12^h35^m$, $12^h46^m$, $12^h57^m$, $13^h08^m$, $13^h19^m$, $13^h30^m$, $13^h41^m$, $13^h52^m$, $14^h03^m$, $14^h14^m$, $14^h25^m$.

We matched the 184245 sources in LAS2 and the 332129 sources in LAS4 to the SDSS Data Release 4 PhotoPrimary catalog with a search radius of 2\arcsec.  There were 423214 (151433 and 271781 matches to LAS2 and LAS4, respectively) matches to SDSS and 93160 (32812 and 60348 in LAS2 and LAS4, respectively) UKIDSS sources that were unmatched.  We required that the candidates be classified as 'stellar' or 'probableStar' in the UKIDSS catalog.  In addition, we require that the optically-detected candidates be classified as stellar in the SDSS catalog as well.

The UKIDSS catalog provides magnitudes for objects measured in a series of circular apertures whose diameters range from 1-13\arcsec. All magnitudes receive an aperture correction approximating the total magnitude for a point source.  \citet{Dye06} state that {\tt aperMag3}, the 2\arcsec\ aperture, provides the most accurate total flux estimate.  We therefore use {\tt aperMag3} for the four UKIDSS filters when computing the colors of our candidates.  Since we are searching for point sources, we use the {\tt psfMag} magnitudes for the SDSS-matched objects.  These are the default {\it asinh} magnitudes on the AB photometric system as defined by \citet{Lupton99}.

We expect that the objects for which we are searching will be fainter than the SDSS detection limit.  Therefore, we assign the unmatched UKIDSS sources optical magnitudes of $i=22$, which corresponds to a signal-to-noise ratio of $\sim 1.5$. Since undetected objects can be many magnitudes below a survey's nominal limit, we assigned the $z$-band limit to be 22 magnitudes, which corresponds to $<1\sigma$ detection; this is a liberal cut, but one which assures that no viable candidates would be missed.  We base these signal-to-noise ratio estimates on thorough Monte-Carlo simulations of faint artificial sources placed in and extracted from SDSS $i$- and $z$-band frames.

\subsection{Color Cuts}

Undetected sources in the UKIDSS catalog are listed with a magnitude of {\tt -9.999995E+008}. To compute colors, we assign such objects the $5\sigma$ detection limits provide in \citet{Dye06}: $Y=20.23$, $J=19.52$, $H=18.73$, and $K=18.06$. The {\it asinh} magnitudes utilized by SDSS ensures that a magnitude is always computed from the measured flux, regardless how small (or negative).  We therefore need not assign magnitude limits to these objects.  

For UKIDSS objects with SDSS matches, we keep only those with $J > 17.5$ and imposed the following color cuts\footnote{Using the values directly from their respective catalogs, we present the SDSS-UKIDSS colors as hybrids of the AB and Vega magnitude systems.  We state the magnitude type explicitly in Equations \ref{sdss_match}, \ref{z6} and \ref{z7}, but drop the subscripts thereafter.} :

\begin{equation}
\begin{array}{c}
(i-z)_{AB} \geq 2.5 \\
\mathrm{and} \\
(z_{AB}-J_{Vega}) \leq 2.0 \quad \mathrm{and}\quad  (i_{AB}-Y_{Vega}) \geq 3.0 \quad \mathrm{and} \quad (Y-J)_{Vega} \leq 0.9 \\
\mathrm{or}\\ 
(i-z)_{AB} \geq 3.5 \quad \mathrm{and} \quad (i_{AB}-Y_{Vega}) \geq 3.0 \quad \mathrm{and}\quad  (Y-J)_{Vega} \leq 0.9 \\
\mathrm{or}\\
(z_{AB}-J_{Vega}) \geq 3.0 \quad \mathrm{and} \quad (J-H)_{Vega} \geq 0.2 \quad \mathrm{and} \quad (J-K)_{Vega} \geq 0.6 \\ \label{sdss_match}
\end{array}
\end{equation}

Assuming the unmatched sources were $i$ and $z$-band dropouts we set $z$ and $i$ to 22 magnitudes, apply the lower limits to the UKIDSS magnitudes as for the matched sources and keep only objects with $J\geq 18$ magnitudes. We then imposed the following color cuts:

\begin{center}
\begin{equation}
(z_{AB}-J_{Vega}) \geq 3.0 \quad \mathrm{and}\quad (J-H)_{Vega}\geq 0.2 \quad \mathrm{and}\quad (J-K)_{Vega} \geq 0.6 \label{z6}
\end{equation}
or
\begin{equation}
(z_{AB}-J_{Vega}) \leq 2.0 \quad \mathrm{and}\quad (i_{AB}-Y_{Vega}) \geq 3.0 \quad \mathrm{and}\quad (Y-J)_{Vega} \leq 0.9 \label{z7}
\end{equation}
\end{center}

The selection criteria in Equations \ref{z6} and \ref{z7} concentrate on quasar colors in different redshift regimes, while avoiding L and T dwarfs.   Based on the quasar tracks and brown dwarf colors from \citet{Hewett06}, Equation \ref{z6} selects objects with $5.8 \lesssim z \lesssim 6.6$ using the the $(z-J)$ vs. $(J-H)$ and $(J-K)$ color-color spaces, while Equation \ref{z7} selects quasars with $z > 7$ using their location in the $(i-Y)$ vs. $(Y-J)$ color-color space.  A band of L and T dwarfs exists at $2 \lesssim (z-J) \lesssim 3$, which we avoid at the expense of excluding quasars in the $6.6<z<7$ redshift range in our selection.

After imposing these cuts we had 269 candidates, 2 of which had SDSS detections and 267 which were undetected by SDSS.   We require that the objects with no SDSS counterpart be detected in at least two UKIDSS bands and eliminated single detections, some of which were obvious cross-talk artifacts.  We also examined 8\arcmin $\times$ 8\arcmin\ UKIDSS cutouts in all four bands for these candidates and eliminated 123 spurious candidates caused by image artifacts.  Many of these were cross-talk images or other artifacts associated with nearby bright stars \citep[see \S7 of][]{Dye06}. 

\subsection{SDSS Images}

Next, we extracted image cutouts for the remaining 146 candidates from SDSS in the $r$, $i$, and $z$ bands using a Simple Image Access Protocol (SIAP) service available through the SDSS Virtual Observatory (VO) tools\footnote{http://skyserver.sdss.org/vo/dr6siap/SIAP.asmx}.  The SIAP allows for the retrieval of SDSS images by specifying a position on the sky, an image size, and desired filter.  We coadded all images containing each candidate which, in some cases, increased the number of passes to three or more images, if multiple passes of SDSS are available.  In addition, we stacked each source's $r$- and $i$-band images to assure no flux is detected blueward of $z$-band dropouts and we stacked the $r$-, $i$-, and $z$-band images to look for faint detections in otherwise undetected sources.
 
We examined these images and eliminated objects affected by image artifacts or proximity to very bright stars.  In addition, there were several isolated, well-detected sources that were not cataloged in the PhotoPrimary catalog, but which were flagged as ``secondary'' SDSS sources. Although secondary objects are defined in the SDSS table schema as re-observations of primary objects, these objects did not have a primary entry in the SDSS catalog and were solely secondary  \citep[a similar observation was made by][]{Maddox08} We removed such objects.  

After these eliminations, we are left with 36 candidates for $z>6.5$ quasars, one of these has an SDSS match and 35 are unmatched in SDSS.  Two pairs of candidates are listed in the UKIDSS catalog as separate sources, but are $<$0.5\arcsec\ apart. We treat each of these as a single source, reducing our candidate number to 34.  We list their average positions and fluxes under a single entry in Table \ref{table:candidates}.

\begin{figure}
\plotone{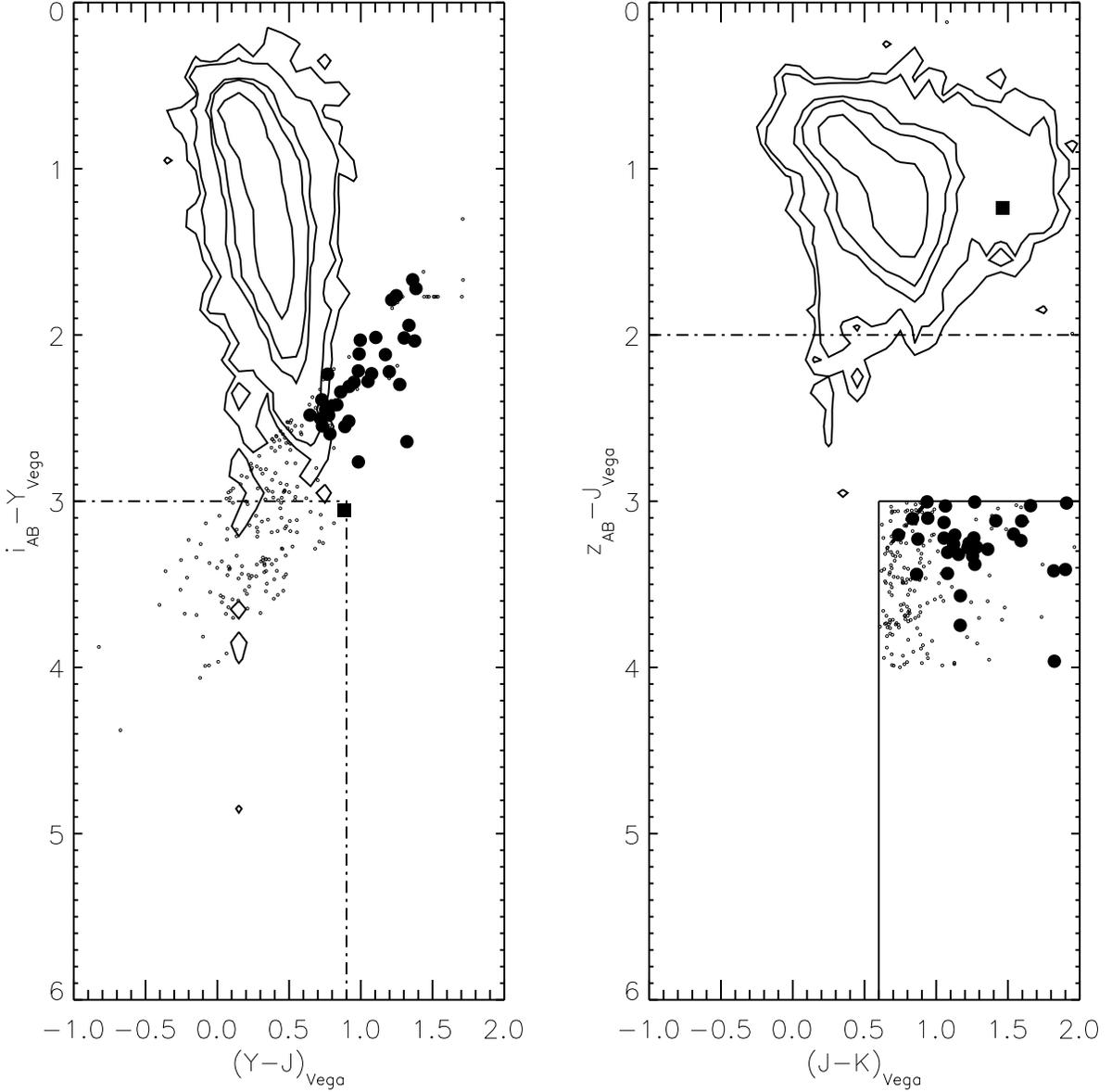}
\caption{Two-color diagrams for our quasar candidates.  On the left, we plot $i-Y$ vs. $Y-J$ and on the right we plot $z-J$ vs. $J-K$.  The contours represent the density of stellar UKIDSS sources {\em with} SDSS counterparts (ie., the stellar locus).  The small black points represent the 267 candidates that obeyed our color selection criteria, including candidates that were subsequently eliminated as various artifacts (as described in the text), while the filled circles represent the 33 objects with no match in the SDSS catalog that passed our visual inspection. The filled square is the only object with a match in the SDSS catalog that passed the color criteria in Equation \ref{sdss_match} and our visual inspection.   The black solid line in the right-hand panel corresponds to the color selection criteria $(z-J)\geq 3.0$ and $(J-K)\geq 0.6$ (Equation \ref{z6}). The dash-dot line in both panels corresponds to the selection criteria $(z-J)\leq 2.0$ and $(i-Y)\geq 3.0$ and $(Y-J\leq 0.9)$ (Equation \ref{z7}). }\label{fig:select_cc}
\end{figure}

We plot in Figure \ref{fig:select_cc} the optical-near-infrared colors of the candidates which did {\em not} have SDSS matches.  On the left, we plot ($i-Y$) vs. ($Y-J$) and on the right we plot ($z-J$) vs. ($J-K$), using the SDSS limits for the $i$- and $z$-band magnitudes. We plot with contours all stellar UKIDSS sources {\em with} SDSS counterparts, representing the stellar locus.  The candidates that obeyed our color criteria are plotted with small black points, while the 34 objects that passed our visual inspection are plotted with large filled symbols; the circles are objects without counterpart in the SDSS catalog and the square is the candidate with a match in the SDSS catalog.  We note that although we use two sets of color cuts for the two redshift regimes (Equations \ref{z6} and \ref{z7}) {\em all} of our optically unmatched candidates are in the higher redshift regime.  Only the object that had a match in SDSS is captured by the lower-redshift color-selection.  

With the SDSS images in hand, we can individually measure the magnitude limit of our candidates in the $i$- and $z$-bands, to better position them in the color-color plots.  We then computed the {\it asinh} magnitude of our source by measuring the sky-subtracted counts in a 4\arcsec-diameter aperture and using the zero-point values provided in the SDSS {\tt Frames} table for each frame \footnote{This process is described in detail in \url{http://www.sdss.org/dr6/algorithms/fluxcal.html}}. We list these in Table \ref{table:candidates}.  When the sky-subtracted flux in the aperture is negative then we list the 1$\sigma$ lower limit magnitude determined from our simulations; $i_\mathrm{lim} = 22.6$ and $z_\mathrm{lim} = 21.1$.  As these measurements show, our estimate of the $i$- and $z$- limits at 22 magnitudes is generally an underestimate for $i$ and an overestimate for $z$, though there is significant variation.

To understand the effects of this variation on our object selection we plot in Figure \ref{fig:calib_cc} the 33 quasar candidates with no SDSS counterpart with square symbols using the now-calibrated SDSS magnitudes in the same color spaces as Figure \ref{fig:select_cc}: $(i-Y)$ vs. $(Y-J)$ on the left and $(z-J)$ vs. $(J-K)$ on the right.  The open squares are spectroscopically confirmed low-mass stars or brown dwarfs, while the filled squares are unidentified sources (see below).  The error bars correspond to the $i$- and $z$-band photometric errors from our aperture photometry measurements.  For comparison, we plot the quasar model tracks with three different continuum slopes from steep to flat, left to right solid lines, respectively,  provided in \citet{Hewett06} and a redshifted elliptical galaxy from \citet{Mannucci01} with open triangles.  We overplot with asterisks and open circles L and T dwarfs and M dwarfs, respectively, using the tables also in \citet{Hewett06}.  The filled circle is a very cool, T8.5 brown dwarf discovered in the UKIDSS First Data Release (DR1) LAS survey \citep{Warren07}.  We plot this object using our imposed SDSS magnitude limits, as it is not detected in SDSS.  Both our color criteria (Equations \ref{z6} and \ref{z7}) reject this object.  In addition, we overplot in the right-hand panel with filled triangles five quasars from \citet{Fan00,Fan01,Fan03} for which $K$-band photometry was available.  

\subsection{PQ Images}

Palomar-Quest (PQ) is a large, digital, synoptic sky survey, conducted at the  Palomar Observatory's 48-inch Samuel Oschin telescope (P48), using the large area QUEST-II camera, a mosaic of 112 CCDs covering  3.6\degr $\times$ 4.6\degr\ \citep{Baltay07,Djorgovski08}.  To date, the survey has covered $\sim 15,500$ deg$^2$ with $-25\degr < \delta < +25\degr$  in up to four filters, either Johnson $U,B,R$ and $I$ or Gunn $g,r,i$, and $z$.  Typical limiting magnitudes in a single pass are $\sim 21$ magnitudes, and in $\sim 6 - 8$ coadded passes under good conditions they reach the depth of SDSS in the redder bands.  The data are taken under almost all reasonable conditions, including non-photometric and poor seeing.  Image full-width at half maxima for stars vary between 2\arcsec\ and $\lesssim 4$\arcsec.

Although the survey data have not yet been fully photometrically calibrated, a mere detection in an $R$- or $r$-band filter is sufficient to eliminate a candidate. We examined co-added PQ images of our 34 candidates in Gunn $r$, $i$ and $z$, and Johnson $R$ and $I$.  One candidate, 132022.09+000211.52, was detected in a Johnson $R$-band image at 13$^h$20$^m$22\fs08 $+$00$\degr$02\arcmin12\farcs53, only 1\farcs02 away.  The co-added PQ image is made up of eight passes.  In addition, the Johnson passband is broader and redder than the SDSS Gunn filter reaching a deeper flux limit in each scan.  The object is visible by eye and a 3\arcsec\ diameter aperture measured the source at a $2\sigma_R$ detection.  132022.09+000211.52 is also detected in the PQ Johnson $I$-band (Figure \ref{fig:pq}) with a $\sim 3.5\sigma_I$ detection, though a possible weaker detection can be seen in the corresponding SDSS image on the right.

\section {Observations}

\subsection{Imaging}

We obtained images for ten of the 34 candidates at the 200 inch Hale Telescope at Palomar Observatory.  Seven fields were observed with the Large Format Camera (LFC)\footnote{\url{http://www.astro.caltech.edu/palomar/200inch/lfc/lfc\_spec.html}} on UT 2006 May 31.  The sky was clear and the seeing was $\sim$1\arcsec.  Since the LFC field of view is $25.3\arcmin \times 24.8\arcmin$ several nearby sources were detected on two of the images. All but one of the fields were observed with a single 300 second integration.  One field, containing 130248.16$-$001917.70, was observed in three dithered frames for 120, 120 and 300 seconds, which we combined to create a final image.

Since any detection in the $r$ or $i$ bands is sufficient to eliminate $z>6.5$ quasar candidate, we used a broad-$RI$ filter with central wavelength of 7670\AA\ and width of 2940\AA\ to collect as many $r$- and $i$-band photons as possible in a relatively short integration time.  All ten sources were detected and therefore eliminated as candidates.  Figure \ref{fig:p200} shows the depth gained by a 300 second exposure image with the P200 (left) over the combined $r$+$i$+$z$ band SDSS image (right).

We reduced the LFC data using the IRAF MSCRED package following standard image data reduction procedures, including an astrometric calibration that was necessary in order to identify our source in the crowded fields, as these images go $\sim 2$ magnitudes deeper than SDSS.  We calibrated the photometry of our $RI$ observations by extracting catalogs with the SExtractor package \citep{Bertin96} and matching the positions to the SDSS catalog.  Choosing $\sim150$ well-detected sources in SDSS that were not saturated in our LFC images, we derived the photometric solution from the $r$- and $i$-band SDSS data.  The $5\sigma$ magnitude limit of these images is $\sim 24\pm0.2$ magnitudes. We measure the magnitude of our candidates in a 2\arcsec\ diameter aperture centered on the position of our candidate and report it in Table \ref{table:candidates}.

Some of these objects were included in the spectroscopic sample, as indicated in the last column of Table \ref{table:candidates}.  All of these were confirmed to be low-mass stars.

\subsection{Spectroscopy}

Spectroscopic followup was conducted in the optical and near-infrared for 27 of our candidates.  Of these we observed all candidates which had no P200 observations, as well as seven of the ten objects with a detection in the $RI$-band P200 images.  

We obtained spectra of 153852.53+053714.43 and 145742.42$+$064058.96 with the LRIS spectrograph \citep{Oke95} on the Keck I telescope on UT 2006 May 21.  We used the red camera with the 600 line mm$^{-1}$ grating blazed at 10000 \AA\ corresponding to a wavelength coverage of $\sim 7900 - 9500$ \AA.  This allows for the detection of Ly$\alpha$ for $z\lesssim 6.8$. We obtained a single 900 second exposure for each object, and although the continuum was detected, the strong fringing could not be removed with a single exposure.  We show the spectra in Figure \ref{fig:opt_spec}. The spectra are similar in shape and are suggestive of late M to early L dwarf spectra. We identify strong stellar absorption features: \ion{Na}{1} $\lambda\lambda$ 8183,9195 \AA\ doublet, a TiO absorption band at 8450\AA, and \ion{Ca}{2} $\lambda$8542 \AA.  We plot in the bottom panel the OH line emission spectrum to identify features from the fringing in the object spectra. The presence of a continuum with no Ly$\alpha$ detection allows us to rule out the possibility of a $z > 6.8$ quasar.

Another 29 spectra for 26 objects were obtained with the near-infrared spectrograph NIRSPEC \citep{McLean98} on the Keck II telescope on Mauna Kea during three observing runs.  We observed during two half-nights on UT 2006 June 11 and June 16, one full night on UT 2007 January 24 and two two-hour blocks on UT 2007 February 08 and February 09, totaling $\sim$2.5 nights.   The weather was clear during all the runs, with the seeing $\lesssim 1\arcsec$.  We observed our candidates in the low-resolution mode using a 0.76\arcsec\ slit for all nights except on UT 2006 June 11, where the $\sim 0\farcs5$ seeing allowed the use of a 0\farcs58 slit.  Most of the objects were observed with 1200 second exposures; a few brighter objects were observed with 600 and 300 second exposures.

We used the NIRSPEC-3 ($J$-band) filter, which spans 1.143 \micron -1.375 \micron, for our initial spectroscopy.  We chose this wavelength window because our expected main contaminant, M, L and T dwarfs, are well-studied in this wavelength regime \citep[e.g.,][]{McLean00} and contain strong neutral potassium (\ion{K}{1}) absorption lines at 1.169,1.177 \micron\ and 1.2432,1.2522 \micron.  A sky-subtracted, dithered pair of well-exposed spectra immediately revealed the \ion{K}{1} absorption features, making for a very efficient observing program.  On all nights, except UT 2006 June 11, we observed at least one A0V star at an airmass representative of the airmasses of our target observations.  We were unable to obtain a calibration spectrum on UT 2006 June 11, and therefore calibrated these data with the A0V star from June 16. 

The data were reduced using the IRAF WMKONSPEC package designed for reducing low-resolution NIRSPEC data\footnote{\url{http://www2.keck.hawaii.edu/inst/nirspec/wmkonspec/index.html}}.  The two-dimensional spectra produced by NIRSPEC are curved and distorted with respect to the detector rows. This software removes the spatial distortions allowing for proper extraction, sky subtraction, and wavelength calibration.  We corrected for telluric absorption in these spectra using the spectrum of the A0V star using the technique outlined in \citet{Vacca03}.

\section {Results}

Given the relatively small area coverage of the EDR, it is not surprising that we do not find any high-redshift quasar detections.  In this exploratory search, we have made deliberately inclusive and liberal color selection of candidates so as to assure that no real quasars would be missed.  It is expected that most of the photometry contaminants in this parameter space would be stars at the bottom of the main sequence or brown dwarfs.  This is indeed what we found in our spectroscopic followup.  

Figure \ref{fig:bd_spec} shows five of our better-detected $J$-band NIRSPEC spectra chosen from the low-mass stars or brown dwarfs found in our survey.  The vertical dotted lines indicate the position of \ion{K}{1} absorption lines at 1.1690, 1.770 \micron\ and 1.2432,1.2522 \micron.  To help identify spectral features that may result from poor sky subtraction or calibration, we plot in the bottom panel of each column the mean OH line emission spectrum and the mean telluric absorption spectrum for our data. This is especially helpful for identifying spectral features in poorly calibrated spectra.

In an effort to classify these stars, following \citet{McLean00} and \citet{McLean03}, we measured the equivalent widths (EWs) of the four \ion{K}{1} absorption lines as well as the strength of the H$_2$O absorption at $\sim 1.34$ \micron.  To determine the line EWs, we fit a multiple component Gaussian plus linear continuum to each doublet. \citet{McLean00} use the 1.33 \micron/1.27 \micron\ ratio to measure the strength of the H$_2$O absorption, because of the strong sky emission at 1.27 \micron, we used the ratio of the mean fluxes in the 1.335-1.345 \micron\ and 1.295-1.305 \micron\ ranges in our measurement.  Table \ref{table:ew} lists the results, as well as the EWs for four comparison objects from \citet{McLean03}.  The EWs of these stars are significantly smaller than those of L and T dwarfs whose EWs rise from 6 \AA\ for early-L to $>10$ \AA\ in late-L and early-T dwarfs \citep{McLean00,McLean03,Cushing05}. M dwarfs have \ion{K}{1} EWs of $\sim 3.5 - 7.5$\AA\ for spectral types ranging from  M6 to M9; the median EWs in our sample is $\sim 3.5$\AA. In addition, H$_2$O absorption in L and T dwarfs is very strong, with 1.33 \micron/1.27 \micron\ $\lesssim 0.55$, while our flux ratios are mostly $>0.7$.  We note, however, that this diagnostic is especially unreliable in our spectra because of the poor calibration of some sources leading to an artificial rise in the continuum beyond 1.33 \micron.

We were unable to classify three of our objects from their $J$-band NIRSPEC spectra.  The \ion{K}{1} absorption lines and the expected water absorption beyond 1.33 \micron\ were not detected in the spectra of 131002.98+000751.88 and 133048.43$-$002316.79 observed on UT 2006 June 11, where no calibration star was available.  We re-observed these objects on UT 2007 Feb 8, calibrating on an A0V star observed less than an hour after the target observations at an airmass difference $<0.01$.  154231.67+053205.29 was observed at the end of the night as the sky was brightening.  Our spectrum has several saturated sky lines which are obvious in the spectrum.  Nevertheless the expected \ion{K}{1} absorption lines at 1.2432,1.2522 \micron\ are in a region of weak sky emission and maximal transmission and are absent from our spectrum.  

It is very unlikely that these objects are high-redshift quasars.  The only redshift range beyond $z=5.7$ (where Ly$\alpha$ moves out of the $i$-band) that is absent of emission line features in the observed $J$-band is $z\simeq 6.1-6.4$. The colors of these objects, however, do not fit this redshift range in either color-color space shown in Figure \ref{fig:calib_cc} and are inconsistent between the two panels for any redshift.  

It is possible that these are compact elliptical galaxies at $z\sim 1.5$.  The seeing in UKIDSS images can be as large as 1\farcs4, corresponding to $\sim 11.5$ kpc at $z=1.5$. In addition, 133048.43$-$002316.79 is classified as 'probableStar' in the UKIDSS LAS catalog.  These objects would require higher resolution imaging to rule them out as galaxies on a morphological basis.

Figure \ref{fig:calib_cc} shows the expected colors of an elliptical galaxy and agrees quite well with our candidates in the $(i-Y)$ vs. $(Y-J)$ color-space.  However, our candidates are too blue in $(J-K)$ and/or too red in $(z-J)$ to fit the expected colors of a $z\sim 1.5$ elliptical galaxy.  It is even less likely that these objects are young, dusty starbursts at similar redshifts.  \citet{Pozzetti00} showed that these objects have even redder $(J-K)$ colors than old ellipticals.  A broader-wavelength ranged spectrum will likely reveal the nature of these objects, though they are highly unlikely to be quasars. 

\section {Discussion}

\begin{figure}
\epsscale{0.8}
\plotone{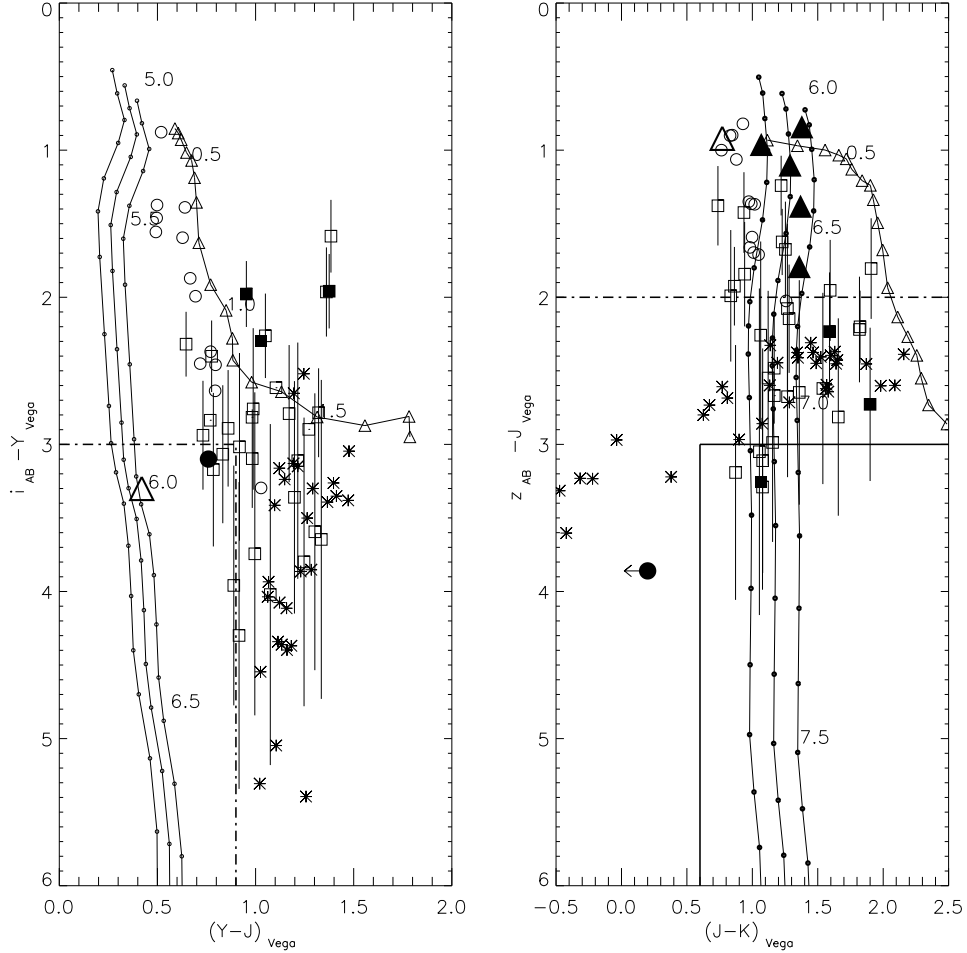}
\caption{Two-color diagrams for the quasar candidates (squares) using the measured {\it asinh} magnitudes on the SDSS system.  The open squares are spectroscopically confirmed low mass stars, while the filled squares are sources with unidentified spectra.  The error bars correspond to the $i$- and $z$-band photometric errors from our aperture photometry measurements. The solid lines show quasar model tracks \citep[from][]{Hewett06} with three different continuum slopes from steep to flat, left to right, respectively.  We plot with connected triangles the colors of redshifted Elliptical galaxy from \citet{Mannucci01} \citep[colors obtained from ][]{Hewett06}.   The asterisks and open circles are L and T dwarfs and M dwarfs, respectively \citep{Hewett06}.  The open triangle is the $z=5.9$ quasar found in the UKIDSS survey by \citet{Venemans07} and the filled triangles are five quasars from \citet{Fan00,Fan01,Fan03} for which $K$-band photometry was available.  The filled circle is the very cool T8.5 dwarf discovered by \citet{Warren07} in the UKIDSS LAS DR1, which is rejected by our color criteria. }\label{fig:calib_cc}
\end{figure}

\begin{figure}
\epsscale{1.0}
\plotone{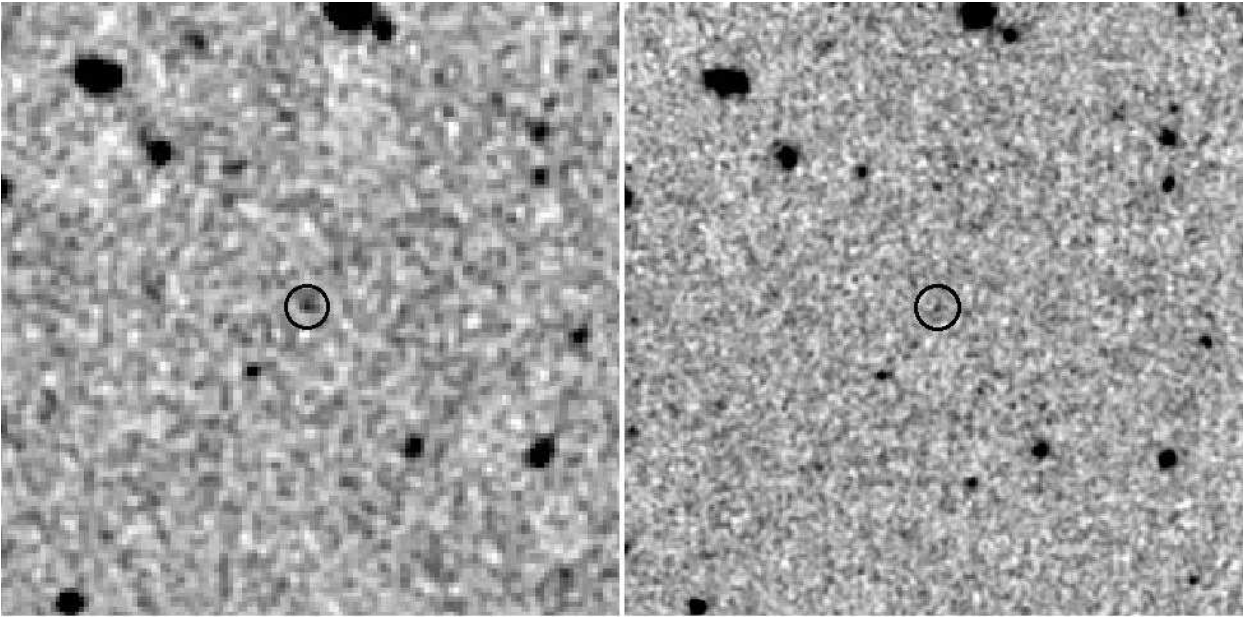}
\caption{2\arcmin$\times$2\arcmin\ images of 132022.09$+$000211.52.  Left -- PQ Johnson $I$-band image which shows a $\sim3.5\sigma$ detection.  Right -- SDSS $i$-band image showing weak detection at the location of the source.}\label{fig:pq}
\end{figure}

The space density of $z\sim 6$ quasars with $M_{1450} < -26.7$ is $6.4\pm2.4 \times 10^{-10}$ Mpc$^{-3}$, assuming $H_0 = 65$ km s$^{-1}$, $\Omega_\Lambda = 0.65$ and $\Omega_m = 0.35$ \citep{Fan04}.  The quasar density evolves with redshift as $\log[\rho(z)] = -0.51z$ \citep{Fan01}, at least out to $z\sim 6$, and we can tentatively extrapolate this trend into the as-yet-unprobed $z>6.5$ regime.  Our selection criteria are sensitive to quasars with $6.0 < z < 6.6$ and $z>7.0$ (see Figure \ref{fig:calib_cc}).  From this, we compute the expected space density of quasars with $M_{1450} < -26.7$ and $7<z<8$ to be $\sim 6\times 10^{-11}$ Mpc$^{-3}$.  The comoving volume of our survey's 27.3 deg$^2$ in that redshift range is $\sim2 \times 10^8$ Mpc$^{3}$.  Although we found no candidates satisfying the low redshift criteria, the space density of quasars with $6.0 < z < 6.6$ is $\sim 1.4\times 10^{-10}$ Mpc$^{-3}$ over a comoving volume of $\sim1.4 \times 10^8$ Mpc$^{3}$.   Therefore, the probability of finding a quasar in the lower-redshift regime is $\sim 0.02$ and in the higher-redshift regime is $\sim 0.01$, given all of the assumptions stated above.  The total probability for finding a quasar in our survey of the UKIDSS EDR is 0.03.   Thus, we did not expect to find any such quasars in this exploratory study, but a detection would have been both interesting and significant.

It is possible that this probability is higher for UKIDSS-selected quasars, since \citet{Venemans07} found, using color selection criteria very similar to those employed here, ULAS J020332.38$+$001229.2, a $z=5.86$ quasar in a subset of the UKIDSS first data release (DR1) covering $\sim$ 4 times larger area and overlapping the SDSS Stripe 82 which reaches $\sim 2$ magnitudes deeper than the single-scan SDSS images. Its absolute magnitude of $M_{1450} = -26.2$ is half a magnitude fainter than the limit for SDSS-selected high-redshift quasars.  This demonstrates that searches for high-redshfit quasars using UKIDSS data and the methodology described here are certainly plausible, even at this early stage.

\citet{Jiang07} found five $z\sim6$ quasars in the SDSS stripe 82 region, including ULAS J020332.38$+$001229.2, and determine the space density for quasars brighter than $M_{1450} < -26.7$ to be $5.0\pm2.1 \times 10^{-9}$ Mpc$^{-3}$, nearly an order of magnitude higher than at $M_{1450} = -26.7$. At these fainter flux limits, the probabilities of finding such quasars rises to 0.15 in the $6.0<z<6.6$ redshift range and 0.09 in the $7.0<z<8.0$ redshift range.  This raises the probability to $\sim 25\%$.  However, despite the increase in sensitivity gained from selecting for quasars in the near-infrared, our small coverage area still makes finding a $z>6.5$ quasar unlikely.

Existing optical datasets that go deeper than SDSS may be useful for reducing stellar contamination, as well as probing deeper into the high-redshift quasar luminosity function, where the density of quasars is higher \citep[e.g.][]{McGreer06,Mahabal05}.

\begin{figure}
\plotone{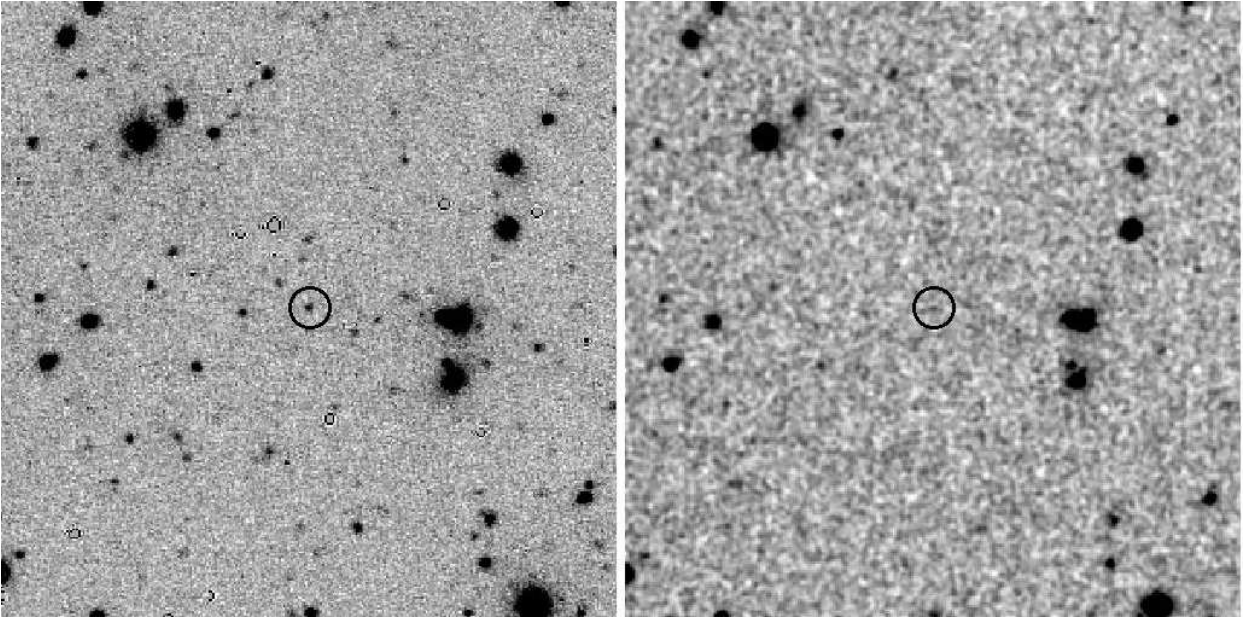}
\caption{2\arcmin$\times$2\arcmin\ images of 153739.52$+$054010.57.  The P200 image (left) shows the depth gained from a 300 second exposure over the combined SDSS $r$+$i$+$z$ image (right).}\label{fig:p200}
\end{figure}

\begin{figure}
\plotone{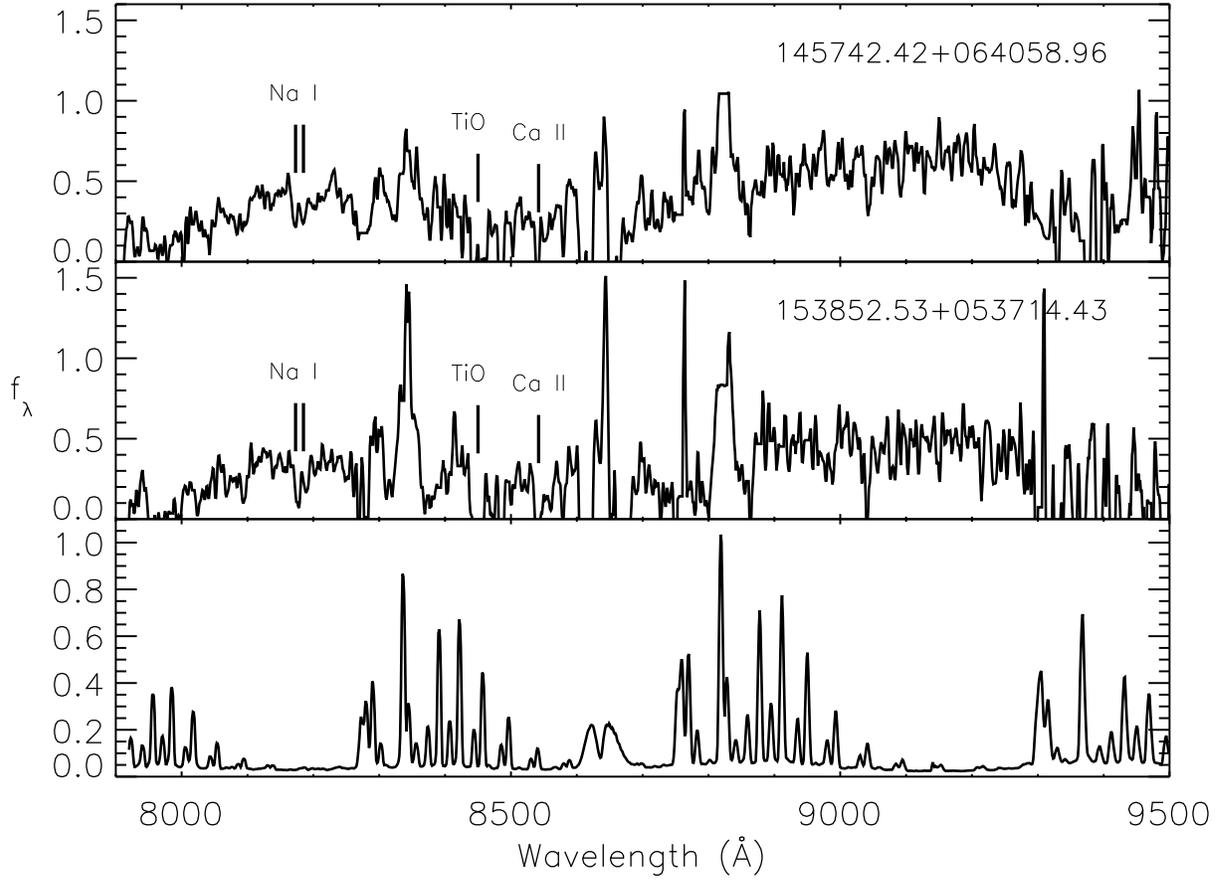}
\caption{LRIS spectra of 153852.53+053714.43 and 145742.42$+$064058.96 showing stellar absorption features, \ion{Na}{1} $\lambda\lambda$ 8183,9195\AA\ doublet, a TiO absorption band at 8450\AA, and \ion{Ca}{2} $\lambda$8542\AA. We plot in the bottom panel the mean OH line emission spectrum for easy identification of sky lines in our spectra.  }\label{fig:opt_spec}
\end{figure}

\begin{figure}
\epsscale{0.5}
\plotone{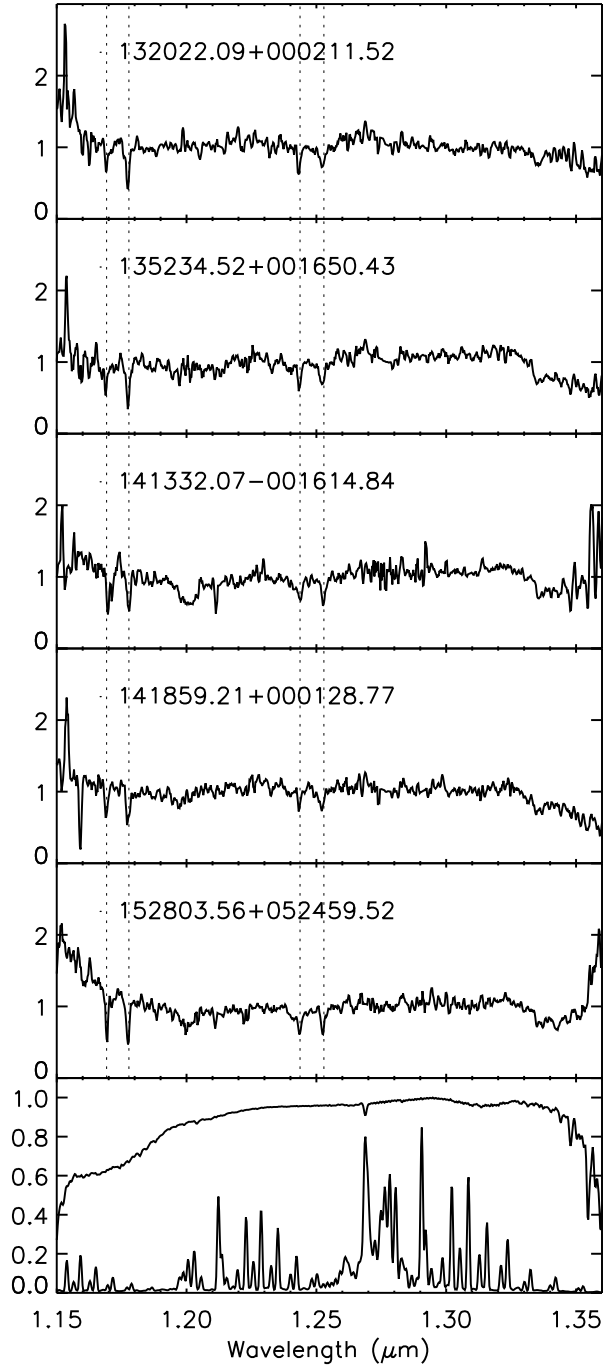}
\caption{A sample of our better-detected $J$-band NIRSPEC spectra of stars found in our survey.  The vertical dotted lines indicate the position of \ion{K}{1} absorption lines at 1.1690,1.770 \micron\ and 1.2432,1.2522 \micron.  We plot in the bottom panel of each column the mean OH line emission spectrum and the mean telluric absorption spectrum for our data.}\label{fig:bd_spec}
\end{figure}

The most important conclusion of this investigation, as already noted by \citet{Hewett06} and \citet{Venemans07}, is that the searches for high-redshift quasars using UKIDSS data will be limited mainly by the depth of the available optical surveys, here SDSS and PQ. Since we are selecting objects that are {\em undetected} in SDSS, the key limiting factor for an expanded highest-redshift quasar search is the depth of the optical imaging.  We have shown, in Figure \ref{fig:p200}, that a 5-10 minute exposure with the P200 telescope can reach 1.5-2 magnitudes deeper than SDSS and will eliminate any low-mass star or brown dwarf candidates.  This is especially important for $z>7$ quasar candidates, where no optical light is expected and a simple detection in a broad-RI filter image will eliminate the candidate from the sample.  It is possible that in order to fully exploit the potential of UKIDSS for discovery of high-redshift quasars, the existing panoramic optical surveys will need to be supplemented by additional wide-field imaging in the far red bands.

An additional area of possible improvement is the computation of updated sets of quasar redshift tracks in these color spaces, which would include both measurements of the optical depth to Ly$\alpha$ and Ly$\beta$ photons as a function of redshift beyond $z\sim 5$ \citep{Fan06}, and a spread of quasar UV-continuum shapes \citep[e.g.,][]{Richards03}.  This can be used to further optimize the color separation of quasar candidate from very red stars and brown dwarfs and provide a better understanding of the high-redshift quasar selection function.  We are currently exploring these issues, and will report on their application in a future paper, although we do not expect a major difference in the resulting color-criteria.\\

We thank Dr. K. Cruz for lending her expertise in low-mass stars to this paper. We are grateful to the staff of W. M. Keck observatory for their assistance during our remote observing runs.  This work was supported in part by the NSF grant AST-0407448, and by the Ajax foundation. This research has made use of data obtained from or software provided by the US National Virtual Observatory, which is sponsored by the National Science Foundation.  F. C., G. M., and D. S. are financially supported by the Swiss National Science Foundation (SNSF). S. G. D. acknowledges the hospitality of EPFL and the Geneva Observatory, where some of this work was performed.


\begin{deluxetable}{cccccccccc}


\tabletypesize{\footnotesize}

\tablewidth{0pt}

\tablecaption{Color-Selected Candidates \label{table:candidates}}

\rotate
\tablehead{\colhead{R.A.} & \colhead{Dec.} & \colhead{$i$} & \colhead{$z$} & \colhead{$Y$} & \colhead{$J$} & \colhead{$H$} & \colhead{$K$} & \colhead{RI\tablenotemark{a}} &\colhead{Notes}\\ 
\colhead{(J2000)} & \colhead{(J2000)} & \colhead{(AB mag)} & \colhead{(AB mag)} & \colhead{(Vega mag)} & \colhead{(Vega mag)} & \colhead{(Vega mag)} & \colhead{(Vega mag)} & \colhead{(AB mag)} & \colhead{} } 

\startdata
 12 59 49.89 & $+$00 01 13.94 &    22.50$\pm$0.42 &    21.23$\pm$0.59  &   19.59 &  18.83 &  18.16 &  17.49& & \tablenotemark{b} \tablenotemark{d} \\ 
 13 00 31.74 & $+$00 11 08.63 &    22.61$\pm$0.40 &    20.92$\pm$0.48  &   19.74 &  18.48 &  17.74 &  17.06& & \tablenotemark{b}\\
 13 01 39.74 & $-$00 07 13.70 &    23.47$\pm$0.53 &    20.75$\pm$0.18  &   20.41 &  19.52 &  18.73 &  18.06& & \tablenotemark{e}\\
 13 02 48.16 & $-$00 19 17.70 & $>$22.6           &    20.83$\pm$0.34  &   19.99 &  18.88 &  18.01 &  17.29& 22.44$\pm$0.09 &\tablenotemark{b}\\ 
 13 03 03.54 & $+$00 16 27.66 &    23.32$\pm$0.80 &    20.42$\pm$0.27  &   20.21 &  19.00 &  18.73 &  18.06& & \\
 13 03 48.63 & $+$00 11 45.02 & $>$22.6           &    21.07$\pm$0.58  &   19.76 &  19.00 &  18.09 &  17.73& 22.65$\pm$0.13 &\tablenotemark{b}\\ 
 13 06 33.64 & $+$00 08 44.84 &    22.55$\pm$0.40 &    20.18$\pm$0.27  &   19.66 &  18.80 &  18.39 &  18.06& & \\
 13 09 55.32 & $+$00 25 24.21 &    23.78$\pm$1.04 &    21.86$\pm$0.70  &   19.48 &  18.57 &  18.15 &  17.49& & \tablenotemark{b}\\ 
 13 10 02.98 & $+$00 07 51.88 &    22.30$\pm$0.30 &    21.23$\pm$0.64  &   20.33 &  18.97 &  18.31 &  17.91& & \tablenotemark{b}\\ 
 13 20 22.09 & $+$00 02 11.52 &    21.92$\pm$0.24 &    21.29$\pm$0.59  &   19.52 &  18.74 &  18.22 &  17.62& & \tablenotemark{b}\\ 
 13 28 35.98 & $-$00 22 13.99 &    22.65$\pm$0.47 &    20.37$\pm$0.23  &   19.58 &  18.75 &  18.18 &  17.52& 22.44$\pm$0.11 &\\ 
 13 28 54.53 & $-$00 23 51.32 &  $>$22.6          &    21.42$\pm$0.65  &   19.78 &  18.80 &  17.94 &  17.26& 22.76$\pm$0.31 &\tablenotemark{b}\\ 
 13 30 48.43 & $-$00 23 16.79 &    21.92$\pm$0.25 &    21.32$\pm$0.52  &   19.96 &  18.59 &  17.47 &  16.69& & \tablenotemark{b}\\ 
 13 32 19.39 & $+$00 20 45.43 &    22.39$\pm$0.37 &    19.96$\pm$0.20  &   19.45 &  18.72 &  17.86 &  17.50& & \tablenotemark{b}\\ 
 13 42 51.33 & $+$00 20 42.84 &    21.84$\pm$0.22 &    21.92$\pm$1.11  &   19.52 &  18.87 &  18.20 &  17.82& & \tablenotemark{b}\\ 
 13 44 44.14 & $-$00 23 59.36 &    23.58$\pm$0.94 &    21.67$\pm$0.68  &   19.98 &  18.68 &  18.06 &  17.52& & \tablenotemark{b}\\ 
 13 52 34.52 & $+$00 16 50.43 &    22.14$\pm$0.30 &    20.24$\pm$0.25  &   19.36 &  18.04 &  17.01 &  16.21& & \tablenotemark{b}\\ 
 14 13 32.07 & $-$00 16 14.84 &    24.04$\pm$0.98 &    20.79$\pm$0.34  &   20.24 &  18.99 &  18.10 &  17.08& & \tablenotemark{b}\\ 
 14 18 59.21 & $+$00 01 28.77 &    22.33$\pm$0.34 &    20.74$\pm$0.38  &   19.24 &  18.25 &  17.47 &  17.09& & \tablenotemark{b}\\ 
 14 20 08.94 & $+$00 14 14.71 &    21.86$\pm$0.25 &    20.89$\pm$0.45  &   20.28 &  18.90 &  18.21 &  18.06& & \\
 14 57 42.42 & $+$06 40 58.96 &    21.70$\pm$0.21 &    20.76$\pm$0.33  &   19.67 &  18.70 &  18.13 &  17.69& & \tablenotemark{c}\\
 14 57 45.72 & $+$06 36 29.70 &    23.71$\pm$1.10 &    21.79$\pm$0.67  &   19.97 &  18.97 &  18.34 &  17.32& 22.44$\pm$0.10 \tablenotemark{f}&\tablenotemark{b}\\ 
 15 26 51.74 & $+$05 46 56.25 &    23.79$\pm$1.16 &    21.80$\pm$0.76  &   19.77 &  18.69 &  18.21 &  17.61& & \tablenotemark{b}\\ 
 15 27 32.25 & $+$05 20 40.81 &    22.67$\pm$0.47 &    21.36$\pm$0.76  &   19.88 &  18.71 &  17.98 &  17.35& 23.15$\pm$0.16 &\tablenotemark{b}\\ 
 15 27 57.36 & $+$05 29 14.41 &    21.94$\pm$0.29 &    20.81$\pm$0.48  &   19.52 &  18.79 &  18.21 &  17.70& 21.77$\pm$0.06 &\tablenotemark{d}\\ 
 15 28 03.56 & $+$05 24 59.52 &    21.98$\pm$0.29 &    20.35$\pm$0.33  &   19.72 &  18.67 &  18.08 &  17.42& 22.33$\pm$0.10 &\tablenotemark{b}\\ 
 15 34 22.57 & $+$05 45 00.62 & $>$22.6           & $>$21.1            &   19.70 &  18.43 &  17.79 &  17.26& 22.33$\pm$0.07 &\tablenotemark{b}\\ 
 15 37 39.52 & $+$05 40 10.57 &    23.70$\pm$1.08 &    20.87$\pm$0.37  &   20.06 &  18.72 &  18.16 &  17.44& 22.77$\pm$0.11 & \\ 
 15 38 52.53 & $+$05 37 14.43 &    22.71$\pm$0.64 &    21.96$\pm$0.87  &   19.69 &  18.77 &  18.35 &  17.90& & \tablenotemark{b} \tablenotemark{c}\\ 
 15 42 31.67 & $+$05 32 05.29 &    21.69$\pm$0.22 &    21.00$\pm$0.40  &   19.72 &  18.76 &  18.00 &  17.17& & \tablenotemark{b}\\ 
 15 47 00.58 & $+$05 47 38.03 &    23.41$\pm$0.82 &    20.49$\pm$0.27  &   19.45 &  18.56 &  17.97 &  17.70& & \tablenotemark{b}\\ 
 15 47 21.08 & $+$05 25 49.45 &    22.58$\pm$0.52 &    21.30$\pm$0.55  &   19.41 &  18.62 &  17.85 &  17.35& & \tablenotemark{b}\\ 
 15 47 27.87 & $+$05 24 51.90 &    23.14$\pm$0.79 &    20.80$\pm$0.36  &   19.78 &  18.58 &  17.66 &  16.76& & \tablenotemark{b}\\ 
 15 51 27.70 & $+$05 28 39.07 &    22.65$\pm$0.55 &    20.74$\pm$0.36  &   19.89 &  18.90 &  18.67 &  17.96& & \tablenotemark{b}\\ 
\enddata
\tablenotetext{a}{Objects in this column were imaged at the 200 inch Hale Telescope at the Palomar Observatory with the Large Format Camera (LFC) using a Broad-$RI$ filter, centered at 7670\AA ($\Delta\lambda = 2940$\AA).  Integration times were 300 seconds, except 130248.16$-$001917.7 which was exposed for 540 seconds.}
\tablenotetext{b}{This object has a NIRSPEC spectrum.}
\tablenotetext{c}{This object has an LRIS spectrum.}
\tablenotetext{d}{This object appeared as a pair of candidates separated by $< 0\farcs5$ in the UKIDSS catalog.  We treat it as a single source and list its average position and flux.}
\tablenotetext{e}{This object was selected as a candidates from the list of UKIDSS sources {\em with} SDSS counterparts.  The $i$ and $z$ magnitudes are the {\tt psfMag} and {\tt psfMagErr} entries in the SDSS PhotoPrimary catalog. }
\tablenotetext{f}{The magnitude derived for this source is highly uncertain due to a grazing cosmic ray at the location of the source. }




\end{deluxetable}


\begin{deluxetable}{cccccc}



\tablewidth{0pt}

\tablecaption{\ion{K}{1} Line Equivalent Widths and the Strength of the H$_2$O Absorption at 1.34 \micron \label{table:ew}}


\tablehead{
\colhead{} & \multicolumn{4}{c}{Wavelength} & \colhead{} \\
\cline{2-5} \\
\colhead{Object} & \colhead{1.169 \micron} & \colhead{1.177 \micron} & 
\colhead{1.244 \micron} & \colhead{1.253 \micron} & \colhead{1.34 \micron/1.29 \micron} \\ 
\colhead{} & \colhead{(\AA)} & \colhead{(\AA)} & \colhead{(\AA)} & \colhead{(\AA)} & \colhead{} } 

\startdata
1259$+$0001 & 7.6$\pm$2.2 & 0.8$\pm$0.9 & 2.4$\pm$0.4 & 1.8$\pm$0.5 & 0.90 \\
1300$+$0011 & 9.4$\pm$1.4 & 1.1$\pm$0.5 & 3.8$\pm$0.3 & 3.3$\pm$0.3 & 0.61 \\
1302$-$0019 & 1.7$\pm$0.5 & 3.9$\pm$0.6 & 0.0$\pm$0.4 & 3.7$\pm$0.5 & 0.71 \\
1303$+$0011 & 3.4$\pm$0.4 & 3.7$\pm$0.4 & 4.4$\pm$0.5 & 5.7$\pm$0.4 & 0.88 \\
1309$+$0025 & 3.2$\pm$0.5 & 0.8$\pm$0.2 & 1.8$\pm$0.2 & 2.6$\pm$0.3 & 0.76 \\
1320$+$0002 & 2.9$\pm$0.4 & 1.6$\pm$0.3 & 4.6$\pm$0.3 & 1.2$\pm$0.2 & 0.88 \\
1328$-$0023 & 3.8$\pm$0.6 & 6.7$\pm$0.8 & 4.3$\pm$0.4 & 3.0$\pm$0.4 & 0.36 \\
1332$+$0020 & 4.1$\pm$0.7 & 2.8$\pm$0.5 & 8.6$\pm$0.6 & 1.9$\pm$0.1 & 0.80 \\
1342$+$0020 & 1.7$\pm$0.2 & 1.8$\pm$0.3 & 1.2$\pm$0.2 & 1.9$\pm$0.1 & 1.23 \\
1344$-$0023 & 3.6$\pm$0.6 & 1.6$\pm$0.3 & 1.7$\pm$0.1 & 1.7$\pm$0.2 & 0.69 \\
1352$+$0016 & 3.4$\pm$0.5 & 2.2$\pm$0.4 & 4.3$\pm$0.3 & 1.2$\pm$0.1 & 0.70 \\
1413$-$0016 & 2.8$\pm$0.5 & 3.1$\pm$0.5 & 3.2$\pm$0.3 & 3.3$\pm$0.3 & 0.73 \\
1418$+$0001 & 2.5$\pm$0.5 & 2.1$\pm$0.3 & 3.5$\pm$0.3 & 0.6$\pm$0.1 & 0.74 \\
1457$+$0636 & 2.6$\pm$0.4 & 3.0$\pm$0.4 & 4.5$\pm$0.5 & 3.1$\pm$0.3 & 0.70 \\
1526$+$0546 & 1.7$\pm$0.2 & 2.4$\pm$0.3 & 2.9$\pm$0.3 & 2.8$\pm$0.3 & 0.70 \\
1527$+$0520 & 0.8$\pm$0.5 & 3.3$\pm$0.5 & 6.0$\pm$0.4 & 1.0$\pm$0.2 & 0.81 \\
1528$+$0525 & 3.1$\pm$0.4 & 2.8$\pm$0.4 & 3.8$\pm$0.3 & 1.2$\pm$0.1 & 0.74 \\
1534$+$0545 & 4.9$\pm$0.7 & 7.0$\pm$1.0 & 5.0$\pm$0.3 & 5.4$\pm$0.7 & 0.83 \\
1547$+$0547 & 3.1$\pm$0.6 & 1.6$\pm$0.4 & 3.7$\pm$0.4 & 3.7$\pm$0.3 & 0.70 \\
1547$+$0525 & 2.9$\pm$0.3 & 2.0$\pm$0.2 & 1.8$\pm$0.1 & 2.6$\pm$0.2 & 1.35 \\
1547$+$0524 & 1.0$\pm$0.5 & 2.9$\pm$0.6 & 4.3$\pm$0.3 & 4.7$\pm$0.6 & 0.67 \\
1551$+$0528 & 1.5$\pm$0.3 & 0.6$\pm$0.3 & 2.1$\pm$0.2 & 2.1$\pm$0.2 & 0.81 \\
\hline
M6 -- Gl 283B & 3.2$\pm$0.2 & 4.9$\pm$0.2 & 3.5$\pm$0.2 & 3.2$\pm$0.2 & \nodata \\
M7 -- VB 8\phn\phn\phd &  4.4$\pm$0.2 & 6.4$\pm$0.2 & 4.6$\pm$0.2 & 4.6$\pm$0.2 & \nodata \\
L2 -- Kelu-1\phn\phn &  3.7$\pm$0.9 & 9.0$\pm$0.7 & 6.9$\pm$0.4 & 6.8$\pm$0.4 & 0.55 \\
L8 -- Gl 337C & 5.8$\pm$0.3 & 8.0$\pm$0.3 & 3.2$\pm$0.3 & 4.8$\pm$0.3 & \nodata \\
\enddata
\tablecomments{The last four objects are low-mass stars from \citet{McLean03} whose equivalent widths are provided for comparison.  The 1.34 \micron/1.29 \micron\ ratio listed for Kelu-1 is from \citet{McLean00}}



\end{deluxetable}

\end{document}